\documentclass[journal]{vgtc} 

\usepackage{cite}
\usepackage[bottom]{footmisc}
\usepackage{mathptmx}
\usepackage{graphicx}
\usepackage{siunitx}
\usepackage{times}
\usepackage[T1]{fontenc}
\usepackage[latin1]{inputenc}
\usepackage{enumitem}
\usepackage[table]{xcolor} 
\usepackage{wrapfig}
\usepackage{multirow}
\usepackage{adjustbox}
\usepackage{mwe}
\usepackage{graphbox}
\usepackage{booktabs}
\usepackage{dirtytalk}
\usepackage{soul}
\usepackage{mdframed}
\usepackage{tabularx}
\usepackage{xcolor}
\usepackage{enumitem}
\definecolor{CindySalmon}{RGB}{232, 125, 114}
\definecolor{greenB}{RGB}{77, 175, 74}
\definecolor{purpleF}{RGB}{152,78,163}

\newcommand{\agwe}[1]{\textbf{\textit{\textcolor{greenB}{#1}}}}
\newcommand{\agae}[1]{\textbf{\textit{\textcolor{orange}{#1}}}}
\newcommand{\wgae}[1]{\textbf{\textit{\textcolor{purple}{#1}}}}
\newcommand{\wgwe}[1]{\textbf{\textit{\textcolor{teal}{#1}}}}

\newcommand*\inline[1]{\includegraphics[height=.9em]{#1}
}
\newcommand{\icon}[1]{\inline{#1.pdf}}

\usepackage[bookmarks,backref=true,linkcolor=black]{hyperref} 
\hypersetup{
 pdfauthor = {},
 pdftitle = {},
 pdfsubject = {},
 pdfkeywords = {},
 colorlinks=true,
 linkcolor= black,
 citecolor= black,
 pageanchor=true,
 urlcolor = black,
 plainpages = false,
 linktocpage
}

\onlineid{1376} 
\vgtccategory{Theoretical & Empirical} 
\vgtcinsertpkg 

\title{Visual Arrangements of Bar Charts Influence Comparisons \\in Viewer Takeaways}

\author{Cindy Xiong, Vidya Setlur, Benjamin Bach, Kylie Lin, Eunyee Koh, and Steven Franconeri, \textit{Member, IEEE}}
\authorfooter{ 
\item
Cindy Xiong is with UMass Amherst.
\newline
E-mail: cindy.xiong@cs.umass.edu.
\item
Vidya Setlur is with Tableau Research.
\item
Benjamin Bach is with University of Edinburgh, United Kingdom.
\item
Eunyee Koh is with Adobe Research.
\item
Kylie Lin and Steven Franconeri are with Northwestern University.
}

\shortauthortitle{Xiong \MakeLowercase{\textit{et al.}}: Effect of Visual Arrangement on Chart Take-Aways}

\abstract{Well-designed data visualizations can lead to more powerful and intuitive processing by a viewer. To help a viewer intuitively compare values to quickly generate key takeaways, visualization designers can manipulate how data values are arranged in a chart to afford particular comparisons. Using simple bar charts as a case study, we empirically tested the comparison affordances of four common arrangements: vertically juxtaposed, horizontally juxtaposed, overlaid, and stacked. We asked participants to type out what patterns they perceived in a chart and we coded their takeaways into types of comparisons. In a second study, we asked data visualization design experts to predict which arrangement they would use to afford each type of comparison and found both alignments and mismatches with our findings. These results provide concrete guidelines for how both human designers and automatic chart recommendation systems can make visualizations that help viewers extract the ``right'' takeaway.}


\begin{document}
\maketitle
\section{Introduction}
Well-chosen data visualizations can lead to powerful and intuitive processing by a viewer, both for visual analytics and data storytelling. When poorly chosen, that visualization leaves important patterns opaque, misunderstood, or misrepresented. Designing a good visualization requires multiple forms of expertise, weeks of training, and years of practice. Even after this, designers still require ideation and several critique cycles before creating an effective visualization. Current visualization recommendation systems formalize existing design knowledge into rules that can be processed by a multiple constraint satisfaction algorithm. Tableau and similar products use such rules to decide whether data plotted over time should be shown as lines or over discrete bins as bars. These systems are useful but rely on simple rules that fail to generalize when additional constraints are added, like the intent of the viewer, their graphical literacy level, the patterns being sought, and the relevant patterns in the underlying data. 

One fundamental problem with existing recommenders is that, while they can correctly specify a visualization type, they offer little or no suggestion for how to arrange the data within the visualization. For example, the same data values can be grouped differently by spatial proximity, as shown in Figure \ref{fig:4designs}. These different visual arrangements can lead to different viewer percepts for the same dataset. For example, the vertical or overlaid configuration might emphasize the strong difference for the two bars in the middle, while the stacked bar configuration might emphasize that group 2 has the highest sum.

Through two studies, we generate a new set of design guidelines for  visual arrangements of bar chart values, as a starting point for visualization interfaces intended to help viewers see the `right' story in a dataset -- one that aligns with a designer's goal. We showed people visualizations, asked them to record their takeaways, and categorized them, generating a mapping between different arrangements of values within a visualization and the types of comparisons that viewers are more likely to make.

\textbf{Contributions:} We contribute an empirical study, studying the effect of visual arrangements on visual comparison, establishing a preliminary taxonomy that can be used to categorize the comparisons that people make within visualizations. We compare the results of our study with expert intuitions, generating design implications that could support natural language (NL) interfaces and visualization recommendation tools.

\section{Related Work}

\label{VisAfford}
Design choices, like picking a chart type or deciding whether to highlight a given pattern, can strongly influence how people perceive, interpret, and understand data. \cite{tversky2014visualizing}. 
Showing the same data as a bar graph can make viewers more likely to elicit discrete comparisons (e.g., A is larger than B), while a line graph is more likely to elicit detection of trends or changes over time (e.g., X fluctuates up and down as time passes)\cite{zacks1999bars}. 
Histograms are effective for finding extremes; scatterplots are helpful for analyzing clusters; choropleth maps are effective for making comparisons of approximate values, and treemaps encourage identification of hierarchical structures \cite{lee2016vlat}. Chart types that aggregate data points, such as bar charts, can lead viewers to more likely infer causality from data compared to charts that do not, such as scatterplots \cite{xiong2019illusion}. Charts that show probabilistic outcomes as discrete objects, such as a beeswarm chart, can promote better understanding of uncertainties \cite{kay2016ish, hawley2008impact, Tait2010, garcia2009communicating}. Showing difference benchmarks on bar charts can not only facilitate a wider range of comparison tasks \cite{srinivasan2018s}, but also increase the speed and accuracy of the comparison \cite{nothelfer2019measures}.

\begin{figure}[h!]
 \includegraphics[width=\columnwidth]{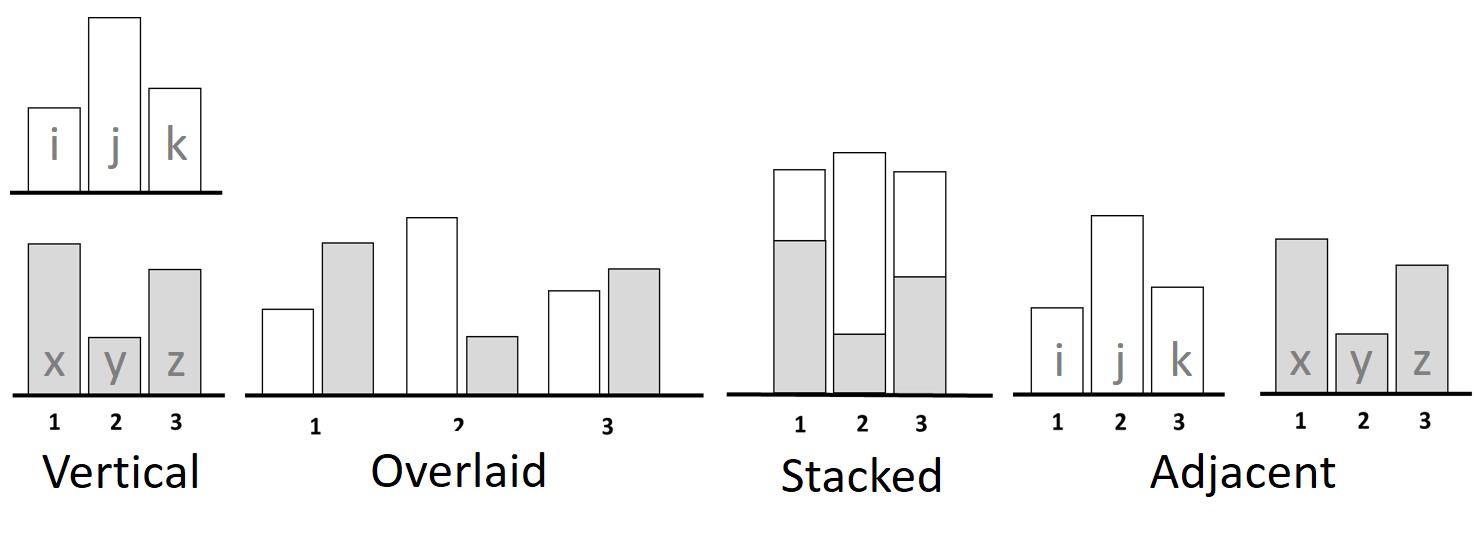}
 \caption{Four spatial arrangements examined in the study.}
 \label{fig:4designs}
 \vspace{-1em}
\end{figure}

Visualizations are often presented in multiples so that analysts can explore different combinations and compare patterns of interest \cite{qu2017keeping}. For example, in interactive visualization dashboards, the spatial arrangement of a visualization can impact decision making, even when the same raw values are displayed \cite{conati2014evaluating}. 
Ondov et al. \cite{ondov2018face} identified four spatial arrangements used to represent multiple views in static visualizations: vertically stacked, adjacent, mirror-symmetric, and overlaid (also referred to as superposed). 
We investigate the effect of four similar spatial arrangements, except that we replaced the mirror-symmetric arrangement, which less commonly used and often for the specific condition of comparing two similar data series \cite{korenjak2008clustering, ondov2018face}, with a more commonly used spatial arrangement: stacked bars, as shown in Figure \ref{fig:4designs}. The adjacent and overlaid arrangements both align bars horizontally, but the adjacent arrangement separates them into multiple x-axes with one group of bars on each. The overlaid arrangement uses a single axis with individual bars of a group next to the corresponding bars from the other group. These four spatial arrangements might encourage different comparisons because they put different values closer to each other. They also differently align values at the same horizontal or vertical positions, which can help viewers compare aligned objects more quickly \cite{matlen2020spatial}.

We hypothesize that participants will more readily compare bars that are visually aligned, and less so the bars that are not. For example, participants might more often compare bar i to bar x, rather than bar i to y, when they view the vertical configuration in Figure \ref{fig:4designs}.

\subsection{Comparisons in Visualization}
\label{CompVis}
Visual comparison has been widely studied, across scenes ~\cite{rensink2002}, scalar fields ~\cite{livingston2011}, and brain connectivity graphs~\cite{alper2013}. It can be a difficult and powerfully capacity-limited cognitive operation. Franconeri~\cite{franconeri2013nature, franconeri2021CurrDirs} discussed multiple cognitive limitations on comparison that should have direct impact on the design displays that facilitate comparisons. For example, objects are easier to compare across translations, relative to transformations of scale or rotation tasks~\cite{xu2015capacity, Larsen1978SizeSI,larsen:1998}.  

Representing comparisons in data visualizations is an important aspect of supporting the user in their analytical workflows. Small multiples make it easier to view objects side-by-side~\cite{ahrens} or examine juxtaposed views through multi-view
coordination~\cite{roberts2007}. Tufte discussed small multiples as an effective  way to use the same graphic to display different slices of a data set for comparison~\cite{tufte_envisioning_1990}. Prior work surveyed a variety of visualization solutions to support comparisons. Graham and Kennedy~\cite{grahamkennedy2007} surveyed a range of visual mechanisms to compare trees, while other surveys consider methods for comparing flow fields~\cite{Post1995}. Gleicher et al.~\cite{Gleicher2011} presented a general taxonomy of visual designs for comparison based on a broad survey of over 100 different comparative information visualization tools. Designs were grouped into three categories: juxtaposition, superposition, and explicit encodings.

Comprehension of visual comparisons is an important aspect of determining their efficacy. Shah and Freedman~\cite{shah2011bar} investigated the effect of format (line vs. bar) on the comprehension of multivariate (three variable) data and found that line and bar chart features have a substantial influence on viewers' interpretations of data. The differences between people's perceptions of bar and line graphs can be explained by differences in the visual chunks formed by the graphs based on Gestalt principles of proximity, similarity, and good continuity. Jardine et al.~\cite{JardineProxies} conducted an empirical evaluation on two comparison tasks -- identify the ``biggest mean'' and ``biggest range'' between two sets of values -- and showed that visual comparisons of largest mean and range are most supported by vertically stacked chart arrangements. More recently, Xiong et al. \cite{xiong2021visual} found that in 2x2 bar charts, people are more likely to group spatially proximate bars together and compare them as a unit, rather than grouping spatially distance bars or comparing bars individually without grouping them.

Based on these, we hypothesize that participants will form visual groups based on spatial proximity (e.g., seeing bar i, j, k in Figure \ref{fig:4designs} as a group, and bar x, y, z, as another group), and make comparisons between bars within a group more often than across different groups.

\subsection{Comparisons in Computational Linguistics}
The ability to establish orderings among objects and make comparisons between them according to the amount or degree to which they possess some property is a basic component of human cognition~\cite{kennedy:2004}. Natural languages reflect
this fact: all languages have syntactic categories (i.e., words in a language which share a common set of characteristics) that express gradable concepts, i.e., expressing explicit orderings between two objects with respect to the degree or amount to which they possess some property (e.g., ``the temperatures in Death Valley are \emph{higher than} in Bangalore in the summer'')~\cite{Sapir1944GradingAS}. Research in computational linguistics has explored the semantics of comparison based on gradable concepts~\cite{CRESSWELL1976261,Hamann2005COMPARINGST,Bierwisch1989TheSO,Klein1980-KLEASF,kennedy:1997,Schwarzchild2002QuantifiersIC}. Bakhshandeh and Allen presented a semantic framework that describes measurement in  comparative morphemes such as `more', `less', `-er'~\cite{bakhshandeh-allen-2015-semantic}.

The semantics of comparatives can be vague as their interpretation depends on the context and the boundaries that make up the definition of the comparative. For the example, ``coffee and doughnuts in the Bay Area are more expensive than in Texas,'' is the statement about whether those items are more expensive \textit{on average}, or whether both items are \textit{individually} more expensive? 
While linguistic vagueness has been explored for comparative expressions along with their semantic variability, little work has been done in determining how best to \emph{visually} represent comparatives based on these variations, especially in the context of visual analysis. Our work explores the types of comparisons readers make and their inherent ambiguities when comparing bar charts in different configurations. 

\subsection{Visualization Recommendation Tools}
Visual analysis tools, such as visualization recommendation (VizRec) systems, can help people gain insights quickly by providing reasonable visualizations. While a detailed review of visualization recommendation (VizRec) systems and techniques is beyond the scope of this paper, it can be found in survey manuscripts such as~\cite{lee2021deconstructing,wu2021survey,zhu2020survey,collins2018guidance}.
Broadly speaking, VizRec systems can be classified based on whether they suggest visual encodings (i.e., encoding recommenders)\cite{Mackinlay1986,Mackinlay2007} or aspects of the data to visualize (i.e., data-based recommenders)~\cite{wongsuphasawat2016towards}.
VisRec systems can provide a specific  recommendation~\cite{datasite,demiralp2017foresight,Lee:2019,scagexplorer,lee2019scattersearch}, but none of these systems focus on how to best provide recommendations specifically for facilitating visual comparison, and offer little or no suggestions for how to arrange the data within the visualization. In this paper, we address this gap in VisRec systems by better understanding how visual arrangements affect the viewers' takeaways during their analysis and the types of comparisons that are made based on these visual arrangements.

\subsection{Natural Language Interfaces for Visual Analysis}
NL interfaces for visualization systems~\cite{thoughtspot,ibmwatson,powerbi} attempt to infer a user's analytical intent and provide a reasonable visualization response. These systems often support a common set of analytical expressions such as grouping of attributes, aggregations, filters, and sorts~\cite{Setlur:IUI,datatone,eviza}. Current NL interfaces however, do not deeply explore how utterances about comparisons ought to be interpreted even though such forms of intent are prevalent~\cite{Setlur:IUI}. In this paper, we explore different ways users express takeaways that compare bars in variants of visual arrangements. The implications of our work also help inform NL interfaces with guidelines towards reasonable visualization responses based on the types of comparisons users specify in their utterances.

\section{Study Motivation and Overview}
We investigate comparison affordances of four spatial arrangements of bar charts by showing crowdsourced participants bar charts and asking them to write sentences describing their most salient takeaways. We analyzed these written takeaways to create a mapping between the visualization arrangements and the takeaways, along with comparisons they tend to elicit. In experiment 2, we compare our data-driven mappings with expert intuitions and generate design guidelines for visualization recommendation systems.

\section{Eliciting Viewer Takeaways in Natural Language}
\begin{figure}[h!]
 \includegraphics[width=3.4in]{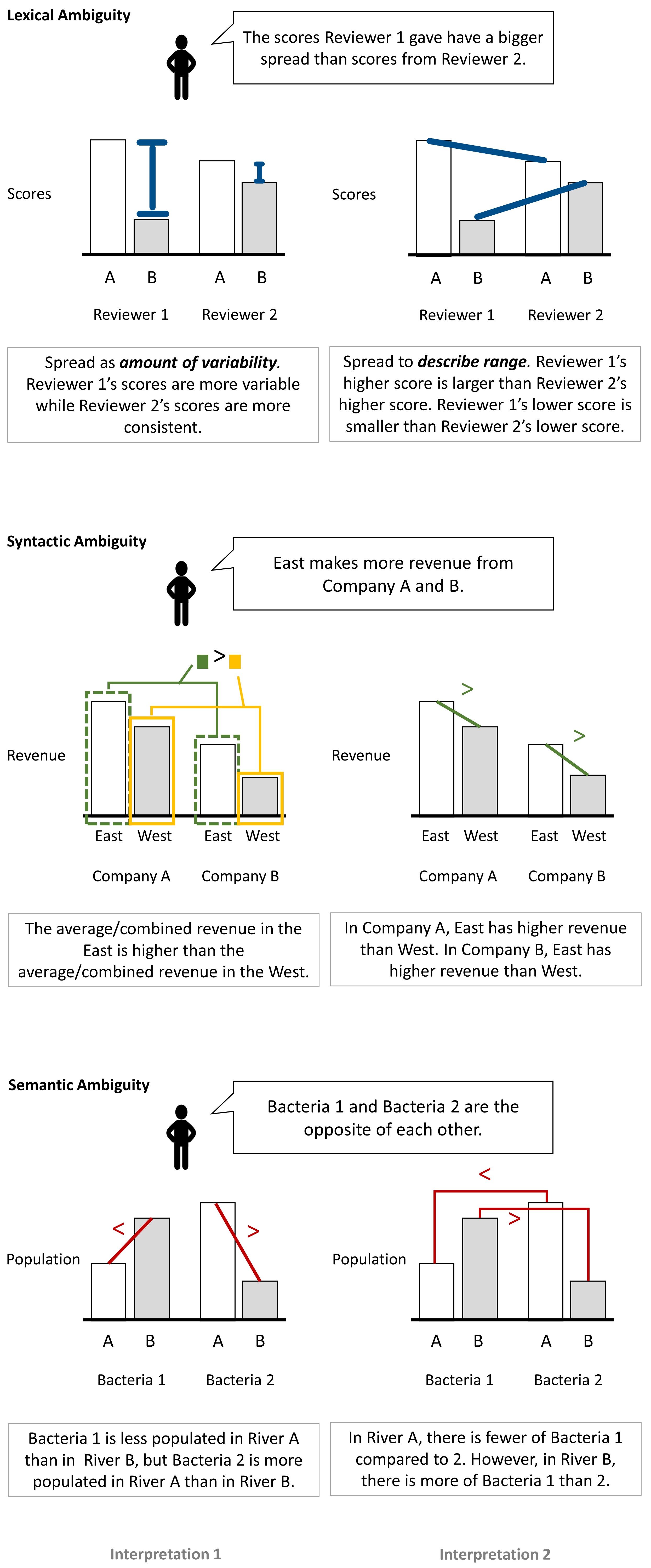}
 \caption{Three linguistic ambiguities for various visual comparisons. Bar charts displayed in the overlaid arrangement.}
 \label{ambiguityDemo}
\end{figure}

One critical challenge in investigating viewer affordances is how to elicit viewer percepts when they interact with visualizations. A dataset can contain many patterns to perceive \cite{xiong2019curse}. For example, looking at the top panel in Figure \ref{ambiguityDemo}, one could notice that both reviewers gave higher scores to $A$ and lower scores to $B$. Alternatively, one could notice that the differences in scores given to $A$ and $B$ is smaller for Reviewer 2 and bigger for Reviewer 1. To communicate what patterns one extracted from these visualizations, the viewer has to generate sentence descriptions of the pattern or relation, such as ``A is greater than B,'' or ``the difference between X and Y is similar to the difference between P and Q.'' In order to examine affordances of different visualization spatial arrangements and to create a mapping between viewer takeaways and the arrangements, we need to interpret and categorize the types of patterns and relations viewers take away from the visualizations. However, we end up facing similar challenges to that of the natural language and linguistics communities \cite{gleich2010ambiguity}. Specifically, the sentences the viewers generate to describe their percepts/takeaways in visualization can be ambiguous. There are three types of ambiguity in natural language: lexical, syntactic, and semantic \cite{gleich2010ambiguity}. Figure \ref{ambiguityDemo} provides an example of each type of ambiguity and how they map to different visual comparisons in the same visualization.

\emph{Lexical ambiguity} represents instances when the same word is used to represent different meanings \cite{kaplan2010lexical}. In our study, we encountered situations where the participants used words such as ``spread,'' which can be interpreted differently depending on their intent. As shown in Figure \ref{ambiguityDemo}, ``spread'' can be interpreted as either the amount of variability in data, or the range of the data as shown in Figure \ref{ambiguityDemo}. 

\emph{Syntactic ambiguity} occurs when there exists multiple ways to parse a sentence. For example, the takeaway ``East makes more revenue from Company A and B'' could be parsed as ``East makes more revenue from (Company A and B),'' or ``East makes more revenue from Company A and (B).'' As shown in Figure \ref{ambiguityDemo}, the viewer could have looked Company A and B holistically and notice that the average or combined values of the East branches is higher than that of the West branches. Alternatively, the viewer could have individually compared pairs of bars, noticing that in Company $A$, the East branch has a higher revenue than the West and that in Company $B$, the East branch has a higher revenue than the West.

\emph{Semantic ambiguity} occurs when multiple meanings can still be assigned to the sentence despite being neither lexically nor syntactically ambiguous. For example, as shown in the bottom panel of Figure \ref{ambiguityDemo}, ``Bacteria 1 and Bacteria 2 are the opposite of each other'' can be mapped to two comparisons. The first could be a comparison between $A$ and $B$ in Bacteria 1 and a comparison between $A$ and $B$ in Bacteria 2, where the former has a smaller than relationship, and the latter has a larger than relationship. The second could be a comparison between Bacteria 1 and 2 in $A$ and another between Bacteria 1 and 2 in $B$.

Since there does not exist natural language processing tools nor existing visual comparison taxonomies to aid our interpretation of chart takeaways, we could not automate the process. We had to manually read every sentence, infer the intent of the participant, and then connect the sentence to a visual pattern in the visualization. The ambiguity in these sentence descriptions can still be vague to even a human interpreter, so we also asked participants to annotate for us which chart component they compared to the best of their abilities. The human interpreter (or researcher, in our case) of these sentences could refer to these drawings and annotations to resolve ambiguities in the sentences. We decided to implement this method after a series of pilot experiments where we failed to comprehensively and accurately capture participant percepts when they viewed visualizations. We describe these failures with the hope that they can inspire future researchers to better capture viewer percepts or takeaways in visualizations.

\noindent\textbf{Attempt 1:} We initially thought that human interpreters of viewer-generated sentences would have little problem resolving the ambiguities in language; unlike machines, we are capable of inferring intention, understanding implicit comparisons, and correcting obvious errors in text. We realized quickly that this was not the case and when a researcher read sentence descriptions as listed in Figure \ref{ambiguityDemo}, they could not reverse engineer the visual patterns the participants extracted. 
 
\noindent\textbf{Attempt 2:} We realized that we needed to ask our participants for more context than just sentence descriptions. If we knew which data values in a visualization they looked at or which pairs of data values they compared, the majority of the ambiguous cases could be resolved. After our participants generated sentence descriptions of the patterns they extracted from a visualization, we asked them to also indicate the data values they compared via a multiple choice task. Consider the chart in the bottom panel of Figure \ref{ambiguityDemo} as an example, the participant would be able to select a subset from the list `Bacteria 1 A', `Bacteria 1 B', `Bacteria 2 A', and 'Bacteria 2 B' to indicate the ones they looked at and compared. However, most comparisons ended up containing the entire set (e.g., a comparison of A1 to B1, and then A2 to B2). In these scenarios, the multiple choice task ends up being uninformative as the participant would select all options in the entire list, because they compared every data value. 
 
\noindent\textbf{Attempt 3:} A sentence typically unfolds as a comparison of two groups in which one group is the `referent' and the other the `target' \cite{clark1972process, gleitman2007give, franconeri2012flexible, roth2012asymmetric}. The target and the referent are connected by a relation. In the sentence ``East makes more revenue than West in Company A,'' the revenue of East A is the target and the revenue of West A is the referent. The relation is  `greater than.' This process applies to both natural language and to visual comparisons across data values in a visualization \cite{shah2011bar, michal2017visual, michal2016visual}. To improve upon Attempt 2, we separated the question where participants indicate which data values they compared into three questions so that they could indicate which values were the target, which were the referent, and the relation between them. 
 
We piloted with $20$ participants, including both crowdsourced workers from Prolific.com \cite{palan2018prolific} and undergraduate students enrolled in a research university and learned that while most people are able to generate sentences describing their percepts, they could not map their comparisons to target, referent, and relations. They especially struggled with implicit comparisons, such as ``there is a decreasing trend from left to right'' and ``West A has the second highest revenue.'' Both cases could be translated into target, referent, and relation in multiple ways. For example, assuming that the participant noticed that the bars became smaller from left to right, the decreasing trend could involve a comparison of the left-most bar to the second left-most bar with the former bigger than the latter. In this case, the target is the left-most bar, the referent is the second left-most bar, and the relation is `bigger than.' Alternatively, the participant could have compared the decreasing trend (the target) to an imagined horizontal line that is not decreasing (the referent). The training process quickly became more complex and its duration became less proportional to its effectiveness. We additionally collected data on participants' confidence as they translated their sentences and observed consistent low confidence in their own translations.
 
\noindent\textbf{Attempt 4:} Inspired by the relation component in the Failure 3, we recognized that mathematical expressions such as `A > B' contain all three elements of target, referent, and relation. Mathematical expressions tend to be far less ambiguous compared to the English language, and writing these simple expressions seems more intuitive than segmenting a sentence into an unfamiliar units. In this attempt, we asked people to write pseudo mathematical expressions to reflect the data values they compared or the pattern they noticed. We provided examples such as `A != C' (A is not equal to C), `A > B > C' (decreasing from A to B to C), and `max = A' (A is the biggest bar) to get people started. After piloting 10 university student participants, we realized that this likely would not scale efficiently to crowd-sourced participants. Participants' expressions varied depending on the type of programming languages they were familiar with. There was little semantic consistencies in how participants used conjunction words like `and', `or', and `but.' For example, some participants used `but' to connect two comparison statements (e.g., A is better than B, but C is worse than D) whereas others used it to represent contrast (e.g., A is better than B, but A is worse than C) or provide context to their comparisons (e.g., they are all the same but A is slightly more). Some sentences were just difficult to be intuitively represented as a mathematical expression, such as ``the population is the same for both rivers, but for different bacteria types.'' 
 
\noindent\textbf{Attempt 5:} This method was a success, but a temporary solution nonetheless. This is the version where we asked participants to write a sentence description and attach a digital drawing annotating the specific patterns they noticed or data values they have compared, as that shown in Figure \ref{drawingComplexityExample}, to clarify the sentence descriptions. What we ended up with was over a thousand sentences and drawings that our later-reported findings are based on. However, this is more of an imperfect, intermediate solution than it is a success -- the method required dozens of hours of manual interpretation from multiple people to ensure that viewer intent is captured accurately and consistently.

\subsection{Lessons Learned}

We share some takeaways from our attempts with future researchers below. First, because there are many patterns to potentially see within our visualization, mapping verbal chart takeaways to visual features is challenging because natural language can be ambiguous. Investigators should try to not rely on sentence descriptions alone to make sense of user intent in the research process. Second, because we do not have tools to automatically interpret viewer takeaways, the research process can become labor intensive, as researchers had to manually decode viewer intents. It will be worthwhile to develop tools that can automate the interpretation of viewer takeaways in the future.

\section{Experiment 1 Crowdsourcing Takeaways}
In Experiment 1, we investigated the comparison affordances of four common arrangements in bar charts: vertically juxtaposed, adjacent, overlaid, and stacked. We asked participants to type out what patterns they perceived and qualitatively coded their takeaways into types of comparisons. We then created a mapping between the visual arrangements and the comparisons they tend to afford. 

\subsection{Participants}
We recruited 76 participants via Prolific.com \cite{palan2018prolific}. They were compensated at nine USD per hour. In order to participate in our study, the workers had to be based in the United States and be fluent in English. After excluding participants who had failed attention checks (e.g., failing to select a specific answer in a multiple choice question to pass the check) or entered illegible/nonsensical response, we ended up with 74 participants ($M_{age} = 25.22$, $SD_{age} = 7.23$, 32 women).  


\begin{figure}[h!]
 \includegraphics[width=\columnwidth]{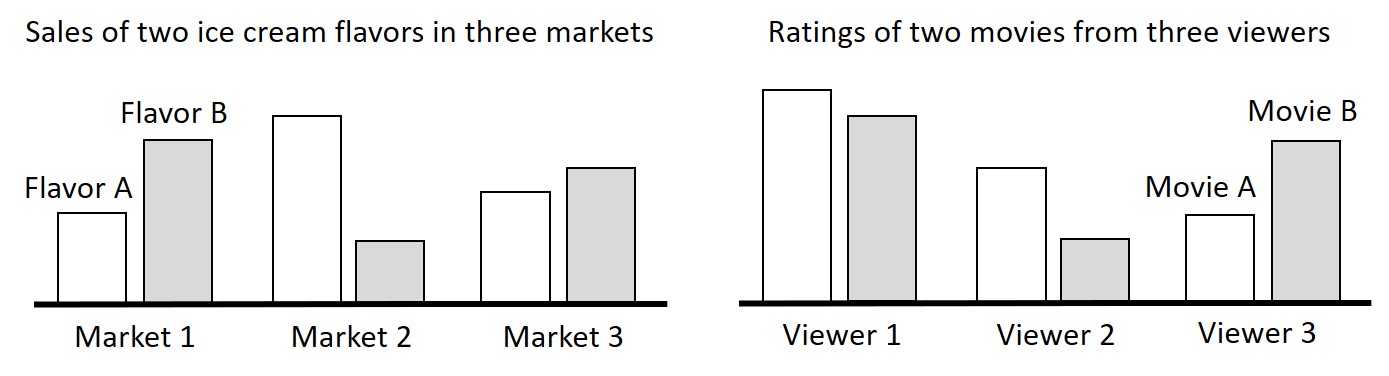}
 \caption{Two datasets used to generate the bar charts, showing the overlaid arrangement as an example.}
 \label{contextExample}
\end{figure}

\begin{figure}[h!]
\centering
 \includegraphics[width=3in]{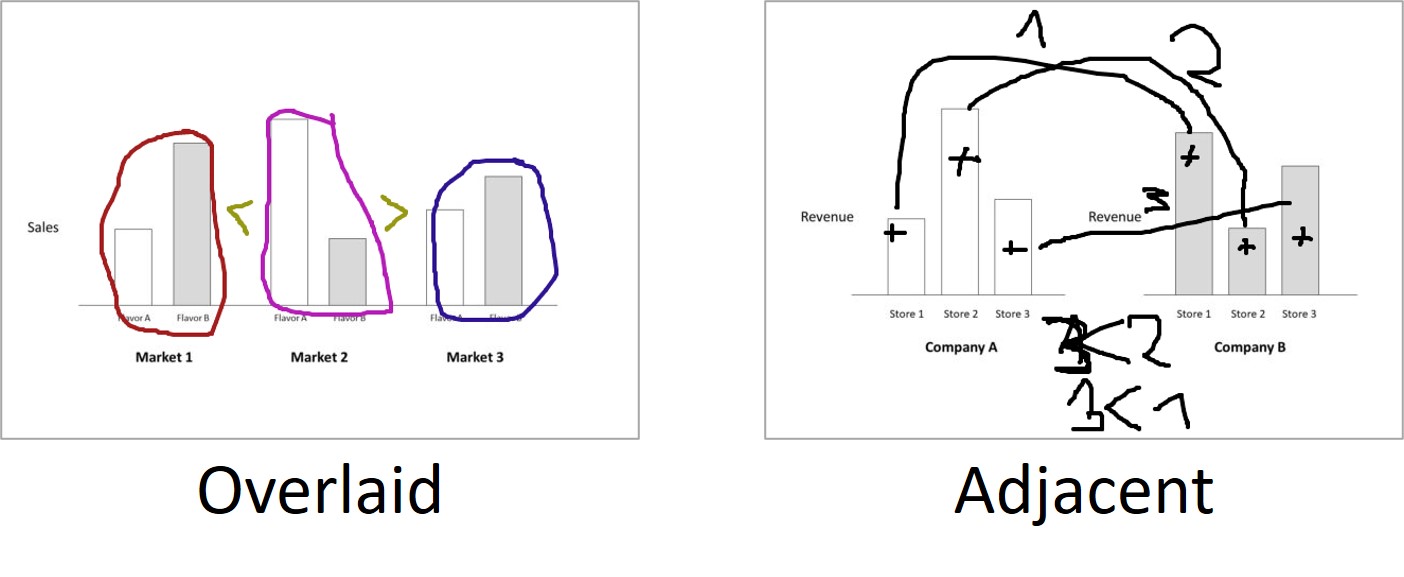}
 \vspace{-2mm}
 \caption{Drawings of a C3 comparison in the overlaid and adjacent charts.}
 \label{drawingComplexityExample}
\end{figure}

\subsection{Methods and Procedure}
We generated two datasets for the four spatial arrangements, creating eight total visualizations. Figure \ref{contextExample} shows the two datasets in the overlaid configuration. Each visualization depicts two groups of three data points. For example, the chart could be showing the sales of two ice cream flavors (flavor A and flavor B) in three different markets (market 1, market 2, and market 3). In our analysis, we will refer to the two groups as `groups' and each of the three data points within each group as `elements.' 

We created a within-subject experiment where each participant viewed all eight of the visualizations and wrote their two main takeaways for each visualization. They were also asked to annotate their takeaways on the bar visualization by drawing circles around the bars they mentioned or using mathematical operators (e.g., $>, <, =$) to represent the patterns they saw, as shown in Figure \ref{drawingComplexityExample}. We examined the sentence takeaways to identify the comparisons participants made upon seeing the visualizations. The takeaways and corresponding drawings can be found in the supplementary materials.

\begin{figure*}[h!]
\centering
 \includegraphics[width = 16cm]{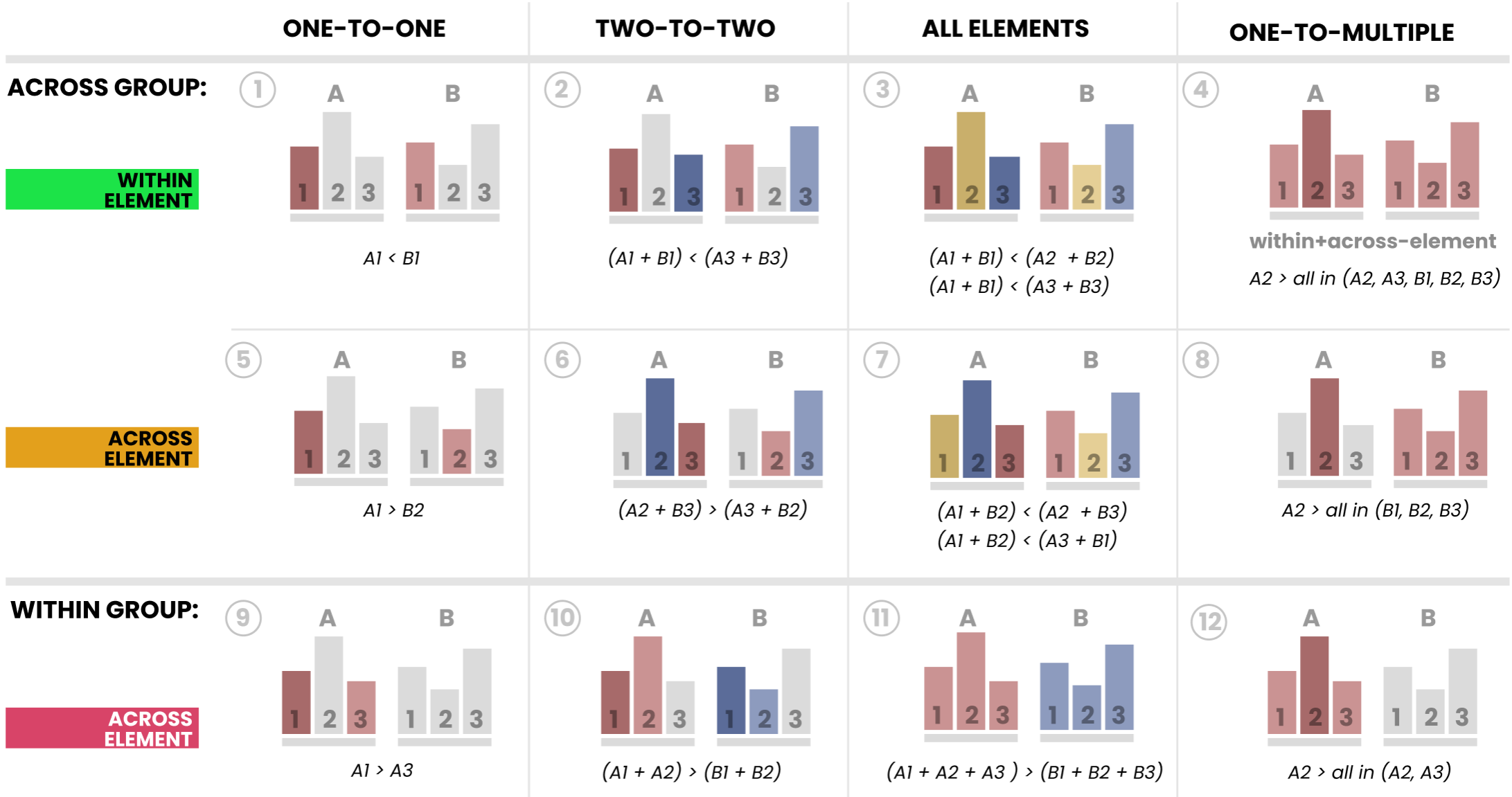}
 \caption{Twelve categories of comparisons in two by three bar charts in the adjacent arrangement.}
 \label{comparisonTypes}
 \vspace{-1em}
\end{figure*}

To distract the participants from noticing similarities in patterns between the charts, we added distractor tasks (e.g., demographic and visual literacy questions) between each visualization and provided each of the eight charts presented with a different context, as shown in Figure \ref{contextExample}. We randomized the order of the charts such that the charts alternated between the two datasets and participants never saw the same spatial arrangements in back-to-back trials. Additionally, we asked at the end of the survey if the participants noticed anything unusual or have any comments regarding the visualizations shown in the survey; six out of the 74 participants mentioned that they noticed similar patterns across the visualizations seen. They mentioned that ``many of the charts were the same, that's why I gave the same answer''  and ``some charts were recycled as the study advanced.'' We decided to keep these participants' data since only a few noticed similarities between our stimuli, and we anticipate them to make our results more conservative (as their answers might differ less among the arrangements).  

\subsection{Comparison Classification and Coding Approach}
We took a top-down approach and identified 12 possible comparisons to generate takeaways from bar charts, as shown in Figure \ref{comparisonTypes}, which we will refer to as C1 through C12. Previously, we mentioned that the charts we have shown participants all depict two groups (A, B) with three elements (1, 2, 3) in each group. A comparison could be made \textit{\underline{across group}}, meaning the viewer compared something in group A to something in group B, or it could be made \textit{\underline{within group}}, meaning the viewer compared something in group A to something else in group A, or compared something in group B to something else in group B. The comparison could also be classified as \textit{\underline{within element}} or \textit{\underline{across element}}. A \textit{\underline{within element}} comparison compares the same elements between two groups, such as comparing element 1 in group A to element 1 in group B. An \textit{\underline{across element}} comparison compares different elements, such as comparing element 1 to element 2. This could be within the same group, meaning element 1 in group A is compared to element 2 in group A, or across different groups, meaning element 1 in group A is compared to element 2 in group B.

A viewer could mix-and-match their comparison operations for groups and elements in four ways: \agwe{across group - within element}, \agae{across group - across element}, \wgae{within group - across element}, and \wgwe{within group- within element}. For an \agwe{across group - within element} comparison, the same element is identified in each group and compared to one another. 
An \agae{across group - across element} comparison means the viewer identifies one element from one group and compares it to another element in a different group. For \wgae{within group - across element} comparisons, the viewer zooms in on one group, and compares different elements within that same group. However, \wgwe{within group - within element}, does not apply in most scenarios since it requires a viewer to compare the same element in the same group, such as comparing element 1 in group A to element 1 in group A. This is just a comparison of a data point to itself, therefore not of interest, and we omit it from our classification system. 

\subsubsection{One-to-One Comparisons (1:1)}
Viewers can compare two individual bars in their takeaways. We refer to them as ``one-to-one'' comparisons, as shown in the leftmost column in Figure \ref{comparisonTypes}, comparison types C1, C5, and C9. For \agwe{across group - within element} operations, one-to-one comparison means that the viewer compares one element in one group to the same element in another group, such as comparing element 1 in group A (which we will refer to as A1) to element 1 in group B (which we will refer to as B1) \icon{1}. For \agae{across group - across element} operations, the viewer compares one element in one group to another element in a different group, such as comparing A1 to B2 \icon{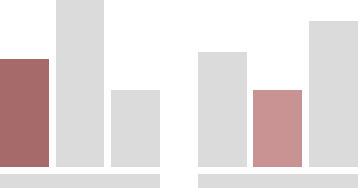}. For \wgae{within group - across element} operations, one-to-one comparison means that the viewer compares one element in one group to another element in the same group, such as comparing A1 to A3 \icon{9}. 

\subsubsection{Two-to-Two Comparisons (2:2)}
Viewers can alternatively visually group together two bars and compare them as a set to another set of visually grouped two bars, which we refer to as ``two-to-two'' comparisons, as shown in the second column in Figure \ref{comparisonTypes}, comparison types C2, C6, and C10. These differ from one-to-one comparisons as the viewer is no longer comparing individual values, but rather comparing the sum/difference of two elements to the sum/difference of two other elements. For example, for an \agwe{across group - within element} two-to-two comparison, the viewer would compare element 1 to element 3 overall \icon{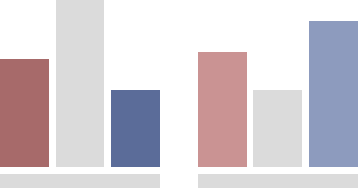}. Using the two ice cream flavor sales across three markets example from before, a comparison of this type will say ``the overall sales in market 1 considering both flavors is lower than the overall sales in market 3.'' For an \agae{across group - across element} two-to-two comparison, the viewer will compare a set of two different elements (one from each group) to another set of two different elements (one from each group) \icon{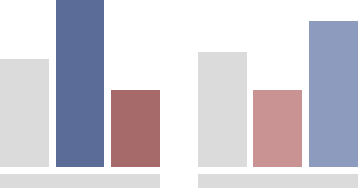}. For a \wgae{within group - across element} two-to-two comparison, the viewer will compare a set of two different elements from the same group to another set of the same two elements from the other group \icon{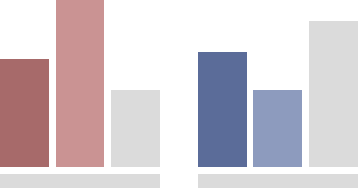}. 

\subsubsection{All Elements (All)}
The previously mentioned two categories involve the viewer comparing a subset of the data in the chart. The viewer can also visually group together a set of bars and compare that set to the remaining data points, which we refer to as ``all element'' comparisons (third column in Figure \ref{comparisonTypes}), comparison types C3, C7, and C11. For this category, an \agwe{across group - within element} comparison would involve the viewer visually grouping the element 1s together and comparing them to the element 2s and 3s \icon{3}. Examples of this type of comparison might include the viewer identifying that the set containing element 1s is overall the smallest compared to element 2s and element 3s (e.g., considering both ice cream flavors, market 1 has the lowest amount of sales compared to market 2 and 3). An \agae{across group - across element} comparison happens when the viewer groups together two different elements, one from each group (e.g., A1 and B2) and compares them to the other pairs of elements \icon{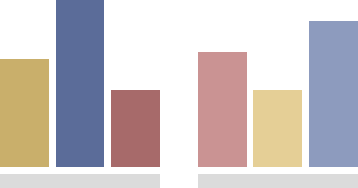}. Since this category requires the elements to not be matching between the two groups, it can seem arbitrary. Our data supports this point as this type of comparison is extremely rare among viewers. Finally, a \wgae{within group - across element} comparison involves the viewer visually grouping together all elements in A and comparing their sum/differences to the sum/differences of all the elements in B as a whole \icon{11}. 

\subsubsection{One-to-Multiple (1:M)}
The last category is ``one-to-multiple'' comparisons, as shown in the fourth column in Figure \ref{comparisonTypes} - types C4, C8, and C12, where participants identify one data point and simultaneously compare it to multiple other bars. People typically do this type of operation when they rank the bars by value (e.g., B3 is the second highest), or when they identify extrema such as maximums or minimums. We refer to the scenario where the viewer picks out one bar and compares it to the rest of the bars as an \agwe{across group - within/across element} comparison because the comparison happened both within the same element (e.g., comparing A2 to B2) and across different elements (e.g., comparing A2 to A1, to A3, etc.) \icon{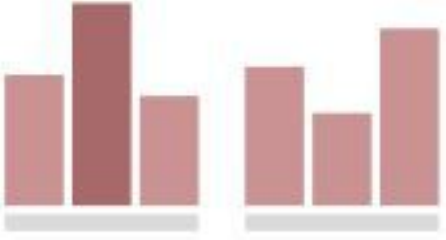}. An \agae{across group - across element} comparison type requires the viewer to identify one element from one group, and comparing it to multiple elements in the other group \icon{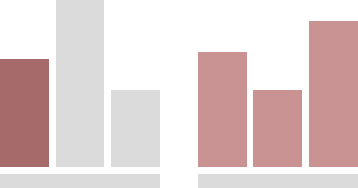}. Lastly, a viewer can identify one element within one group and compare it to all of the other elements in the same group, which we refer to as a \wgae{within group - across element} comparison \icon{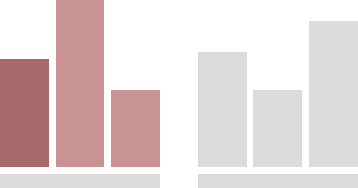}.

\subsubsection{Hypotheses in Context}
Following the hypotheses we proposed in Sections \ref{VisAfford}, we anticipate participants to do more \agwe{across group - within element} comparisons, because they require the viewer to compare the spatially aligned bars, and more \wgae{within group - across element} comparisons, because they require the viewer to group bars together and compare the bars within the same group. Additionally, they might do more 1:1 comparisons than everything else, as 1:1 comparisons have the most straightforward alignment. However, if participants are actually more likely to group spatially proximate bars to compare them, we should see an interaction between comparison type and chart arrangement. For bar charts in the vertical and stacked arrangements, we should see more C1, C2, and C3 because they are vertically aligned \textit{and} spatially proximate, which makes them intuitive to compare, or C9 and C12 because they involve comparing bars within a spatially proximate group. Following a similar logic, C9 and C12 might also be often compared in the adjacent arrangement. For the overlaid arrangement, because the bars are grouped together by element pairs (1, 2, 3), we expect to see fewer of C9 and C12, and more of C2 and C3, as viewers will likely group the element pairs together to compare one pair with another pair.

\subsection{Making Sense of Participant Takeaways}
To make sense of participant takeaways, we analyzed their written description in conjunction with their drawings, as shown in Figure \ref{drawingComplexityExample}. These drawings are especially helpful when we were not sure what elements the participants compared exactly based solely on their written descriptions. Most takeaways involved a comparison between two chart elements, so we identified what they compared in each takeaway and what relationship described that comparison, mapping that comparison to one of our twelve categories. For example, for the takeaway ``it looks like flavor B sold more than flavor A in market 2'', the two chart elements compared would be B2 (flavor B in market 2) and A2 (flavor A in market 2), and the relation is `greater than.' Since this is a comparison of one individual bar to another individual bar, this would be a one-to-one comparison. Additionally, since the element is fixed (market 2 for both) and the group is changing, this would be an \agwe{across group - within element} comparison (type C1). Two authors participated in this qualitative coding process and double coded all responses. They agreed 89.5\% of the time in their ratings, with a high inter-rater reliability Kappa value of 0.867 ($z = 72.3$, $p < 0.001$). Disagreements were resolved through discussion.

Some participants would make multiple accounts of the same type of comparison upon seeing one chart, such as making two instances of across group - within element one-to-one comparisons (see 1 in Figure \ref{comparisonTypes} \icon{1}). For example, one participant wrote ``I noticed that in market 1, flavor B sold more than flavor A, and in market 2, flavor A sold more than flavor B.'' This participant compared B1 to A1, as well as compared B2 to A2. In situations like this, we removed the duplicate and counted this participant as having done a one-to-one \agwe{across group - within element} comparison. 

We also noticed that some participants mentioned conjunction comparisons in their takeaways, touching on two comparison categories, such as saying ``flavor B sold more than flavor A in market 1, but flavor A in market 1 sold more than flavor A in market 3.'' In this example, the first part of this comparison is a one-to-one \agwe{across group - within element comparison} (C1 \icon{1}), and the second part of this comparison is a one-to-one \wgae{within group - across element} comparison (C9 \icon{9}). We recognize that conjunctions could also be used to join two takeaways as a higher order comparison (e.g., a comparison of the result of two previous comparisons). However, related work has shown that these cases are relatively rare as higher order comparisons are complex and people seldom make them \cite{xiong2021visual, shah2011bar}. Thus for the present experiment, we break up conjunctions, which cover different types of comparisons, and treat them as separate takeaways from the same participant.

\subsection{Comparison Overview}
\begin{figure}[h!]
 \includegraphics[width=\columnwidth]{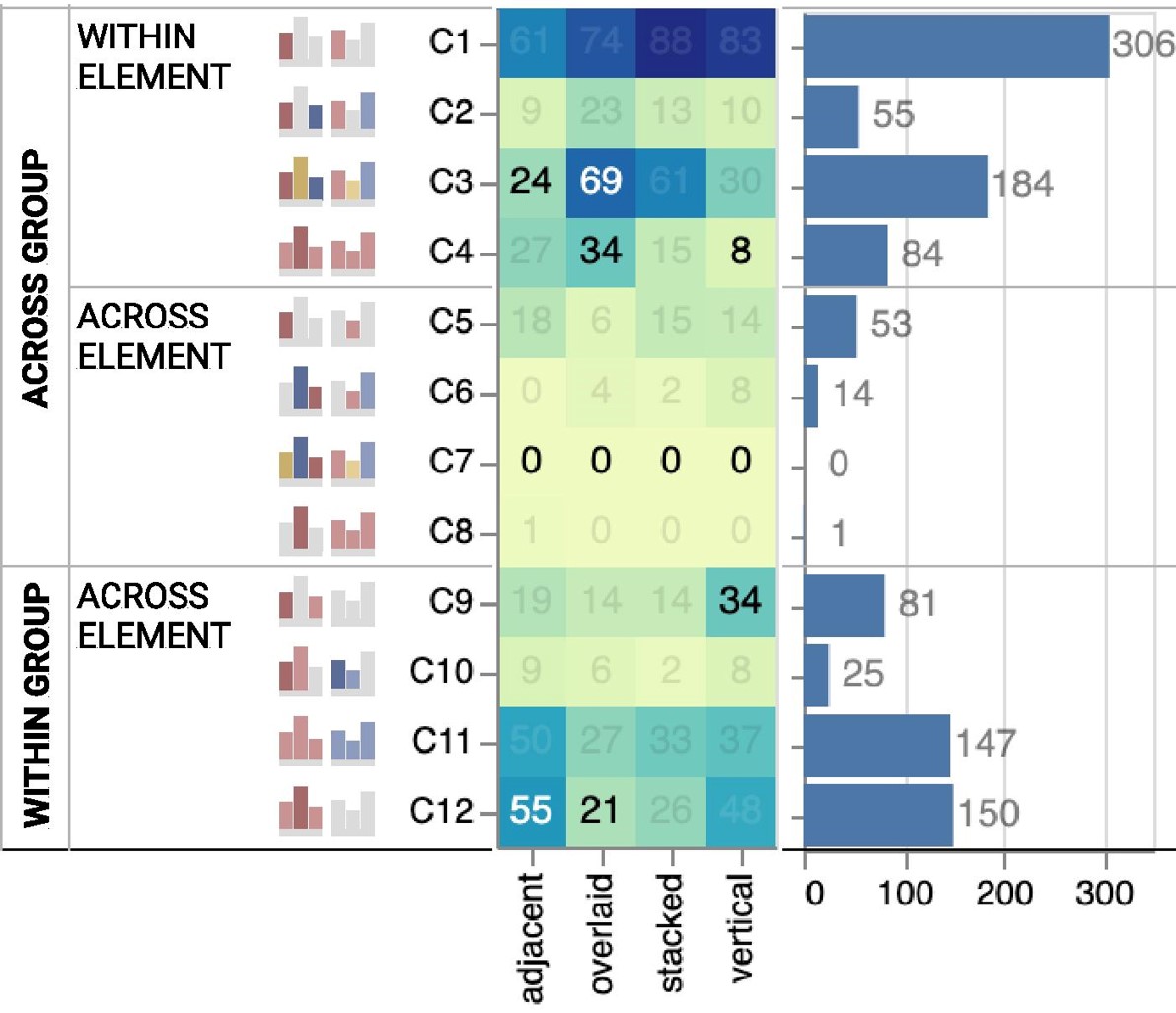}
 \centering
 \caption{Heatmap showing occurrences of categories for every comparison type as well as in total across all types (bar chart). Opaque values in the heatmap indicate significant values. Bar chart icons shown on the left are presented in the adjacent arrangement.}
 \label{heatmap_12categories}
\end{figure}

We collected a total of 584 chart takeaways with 53.09\% of them being conjunction comparisons, with a total of 1100 comparisons collected from participants. Figure \ref{heatmap_12categories} and Table \ref{exp1_overview} summarize the number of comparisons participants made for each category and shows the frequencies of the 12 types of comparisons across the four visualization arrangements.

\begin{table}[h!]
\centering
\scriptsize
\setlength\tabcolsep{5pt}
\begin{tabularx}{\linewidth}{lllp{5.2cm}}
	\toprule
	 & Type & $\%$ & \textbf{Examples}\\ \hline
    \textbf{1} & C1 & 27.82 
    & compare one element in a group to the same element in the other group \\ \arrayrulecolor{lightgray}
    \hline
    \textbf{2} & C3 & 16.73 & compare one element to the other two elements considering both groups \\\hline
    \textbf{3 (tie)} & C11 & 13.36 & compare the group A as a whole group B as a whole\\ \hline
    \textbf{3 (tie)} & C12 & 13.36 & compare one element in one group to the other elements in the same group \\ \hline
    \textbf{5} & C4 & 7.64 & select one data point and compare it to all the other data points \\ \hline
    \textbf{6} & C9 & 7.36  & compare one element in one group to another element in the same group\\ \hline
    \textbf{7} & C2 & 5.00 & compare one element to one other element considering both groups \\ 
\end{tabularx}
\vspace{2pt}
\caption{Overview ranking the most frequently made comparisons from Experiment 1 with examples.}
\label{exp1_overview}
\vspace{-2mm}
\end{table}

Participants most frequently made the \emph{one-to-one \agwe{across group - within element}} comparisons (C1 in Figure \ref{comparisonTypes}). Using the ice cream flavor sales across three markets example, where the two groups are flavors A and B and the three elements are markets 1, 2, and 3, participants most often wrote takeaways saying ``I noticed that in market 1, ice cream flavor A sold less than flavor B.'' 

The second most frequently made comparison was \emph{all elements \agwe{across group - within element}} (C3). For example, ``the total sales in market 1 is smaller than the total sales in market 2, and it's also smaller than the total sales in market 3.''

The third and fourth (tie) most frequently made comparisons were \emph{all elements \wgae{within group - across element}} (C11), such as saying ``overall, flavor A sold more than flavor B,''
and \emph{one-to-multiple \wgae{within group - across element}} (C12), such as saying ``for flavor A, market 2 sold more ice cream than market 1 and market 3.''

The fifth most frequently made comparison was \emph{one-to-multiple \agwe{across group - within/across element}} (C4). This type of comparison mostly involved identification of the maximum or the minimum points, or a ranking of data values. For example, ``flavor B in market 2 is the least sold flavor, considering all the flavors and all the markets.''

The sixth most frequently made comparison was \emph{one-to-one \wgae{within group - across element}} (C9). For example, one participant said ``For flavor A, market 1 sold more than market 3 did.''

The seventh most frequently made comparison was the \emph{two-to-two \agwe{across group - within element}} comparison (C2). This is similar to the all element across group - within element comparison (C3), except that the participant only compared one element to one other element. For example, ``market 1 has better business than market 3.''

Participants very rarely did a comparison of the remaining five categories (C5, C6, C7, C8, C10), making up less than 5\% of all comparisons each. No participants made \emph{all elements \agae{across group - across element}} comparisons (C7), which means no one visually grouped non-matching elements from the two groups and compared them. Only one participant made a \emph{one-to-multiple \agae{across group - across element}} comparison (C8), where they compared the biggest bar in group A to all of the bars in group B. The \emph{one-to-one \agae{across group - across element}} comparisons (C5) that participants made may seem arbitrary, but they were always comparing two bars of similar sizes together. This finding supports our hypothesis from Section \ref{VisAfford}, suggesting that most comparisons centered around spatially aligned elements in bar charts.

\begin{figure}[h!]
 \includegraphics[width=\columnwidth]{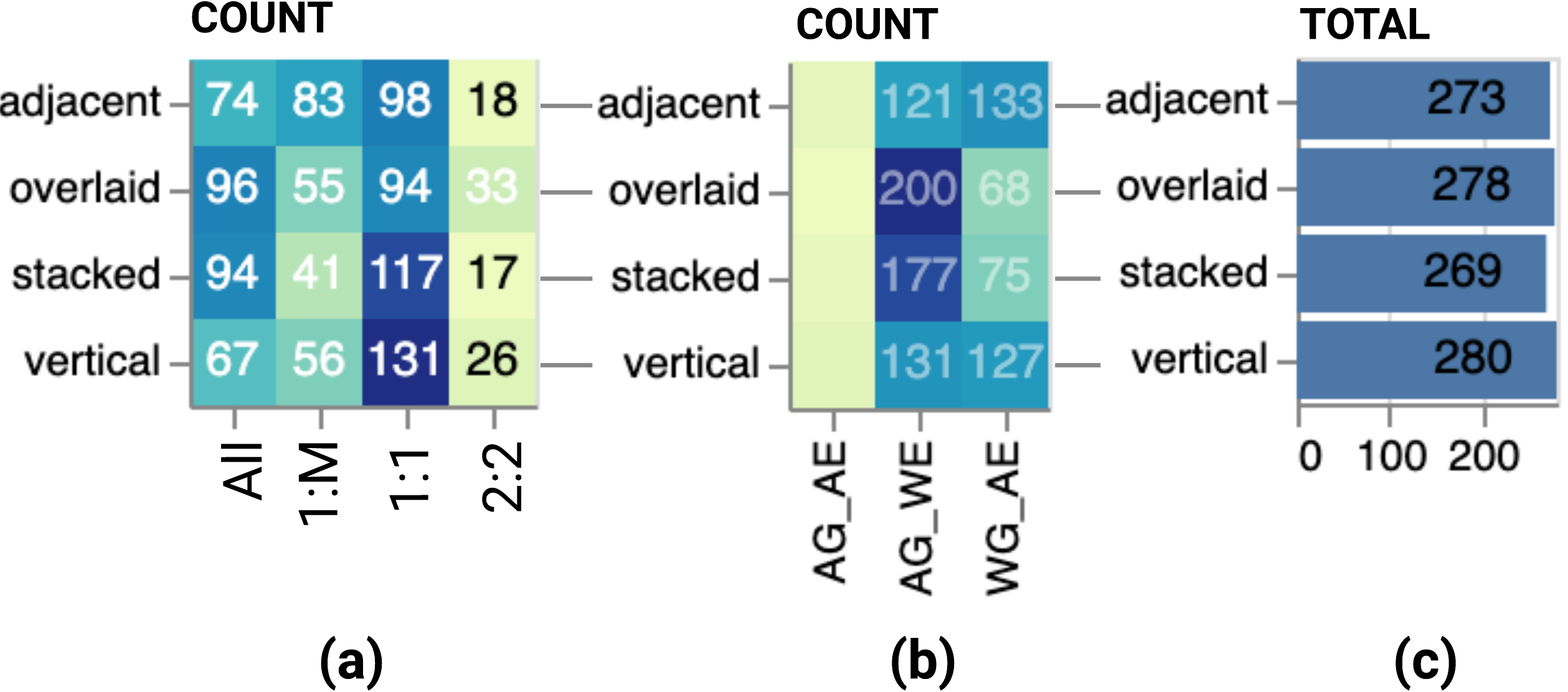}
  \centering
 \caption{Heatmaps showing (a) the total counts of comparison types for each group (1:M, 1:1, 2:2), (b) per groupings (AG, WG, AE, WE) [no significant counts], and (c) the total count of observations across techniques. Opaque values in the heatmap indicate significant values.}
 \label{exp1_groupResults}
\end{figure}

\begin{table*}[ht]
\centering
    \scriptsize
    \caption{Summary results from Experiment 1 (crowdsourced comparisons) compared to summary results from Experiment 2 (expert intuitions). Bar charts in the right-most column show expert preferences for the four arrangements: adjacent (a), overlaid (o), stacked (s), and vertical (v).}
    \vspace{3mm}
    \begin{tabular}{lp{4cm}lp{4cm}lll}
   \textbf{Type} & \textbf{Comparison} & \textbf{}& \textbf{Comparison Goal} & \textbf{Crowdworkers} & \textbf{Experts} & \textbf{} \\ 
   \hline 
    C1 
    & compare the same element across two different groups
    & \includegraphics[align = t,scale = 0.3]{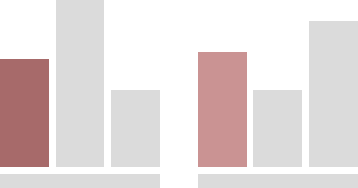}
    & compare the sales revenue in market 3 from ice cream flavor A to the revenue in market 3 from ice cream flavor B
    & all the same & vertical 
    & \includegraphics[align = t,scale = 0.23]{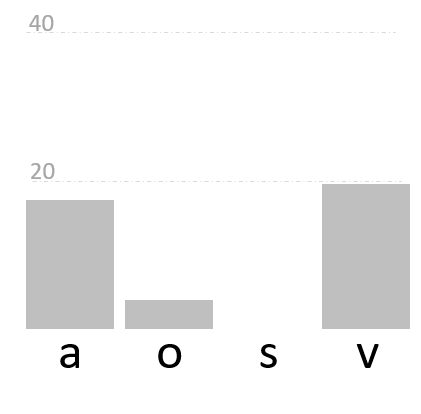}  \\  \hline
    
    C3 
    & compare one element to other elements, considering both groups
    & \includegraphics[align = t,scale = 0.3]{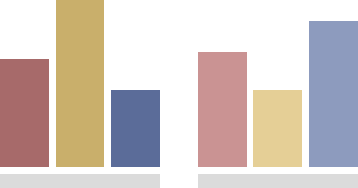}
    & compare the average sales revenue in market 1 to the average sales revenue in market 2, across both ice cream flavors
    & overlaid  & stacked 
    & \includegraphics[align = t,scale = 0.23]{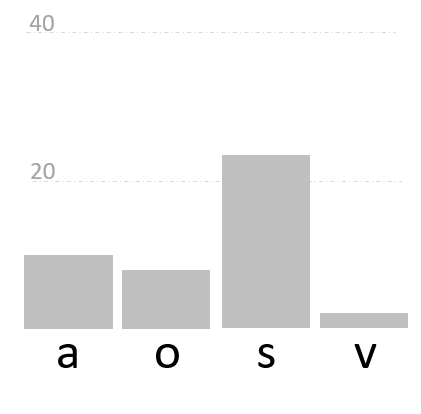} \\  \hline
    
    C4 
    & compare one bar to all other bars in the chart, which is often a superlative comparison
    & \includegraphics[align = t,scale = 0.25]{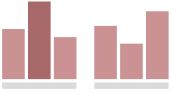}
    & identify the lowest single sales revenue rating among all six revenues
    & overlaid  & overlaid 
    & \includegraphics[align = t,scale = 0.23]{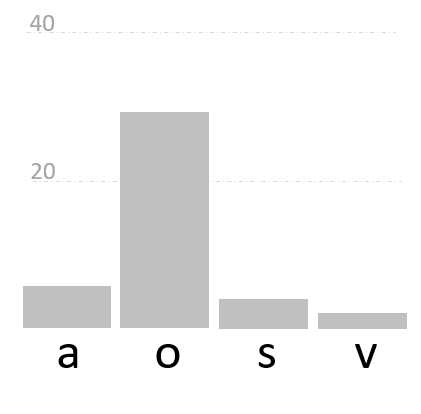} \\  \hline
    
    C9 
    & compare one element in one group to another element in the same group 
    & \includegraphics[align = t,scale = 0.3]{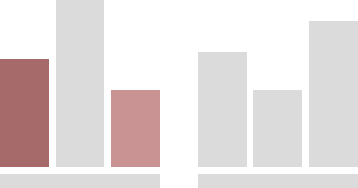}
    & compare the sales revenue of ice cream flavor B in market 1 to the sales revenue of ice cream flavor B in market 3
    & vertical  & overlaid 
    & \includegraphics[align = t,scale = 0.23]{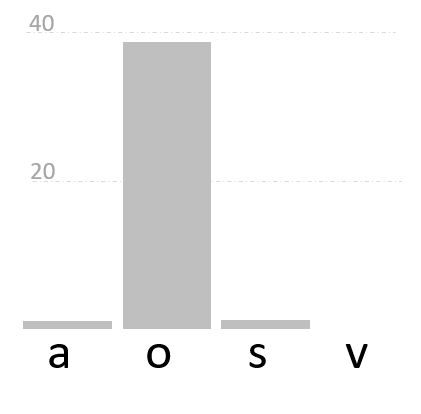} \\  \hline

    C11 
    & compare one group to another holistically
    & \includegraphics[align = t,scale = 0.3]{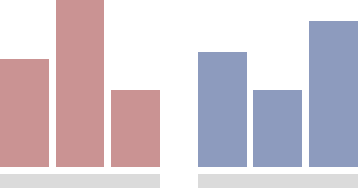}
    & compare the overall ice cream sales in market A to the overall ice cream sales in market B
    & all the same  & all the same 
    & \includegraphics[align = t,scale = 0.23]{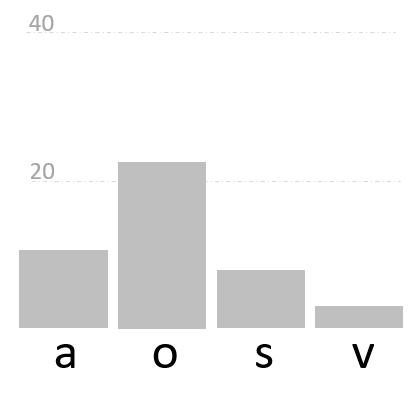} \\ 
\hline
\end{tabular}
\label{exp2design}
\vspace{-1em}
\end{table*}

\subsection{Visual Arrangements and Number of Bars Compared}
We analyzed whether different visual arrangements impacted whether a viewer would do one-to-one (1:1), two-to-two (2:2), all elements (All), or one-to-multiple (1:M) comparisons (the four columns in Figure \ref{comparisonTypes}). As shown in Figure \ref{exp1_groupResults}, as we expected, most participants made 1:1 comparisons (40.0\%), followed by All comparisons (30.1\%) and 1:M comparisons (21.4\%), and the least 2:2 comparisons (8.5\%).

Considering the specific visual arrangements, another Chi-Square analysis found that the only significant relationship is that the adjacent arrangement affords more one-to-multiple (1:M) comparisons, while the stacked arrangement affords fewer such comparisons (${\chi}^2$ = 38.405, $p < 0.001$). This means visual arrangement in general does not affect how many bars people compare in a chart.

\subsection{Across/Within Group and Element Comparisons}
We conducted a Chi-Square analysis and, as shown in Figure \ref{exp1_groupResults}, found that participants were significantly more likely to make \agwe{across group - within element}, and \wgae{within group - across element} comparisons (${\chi}^2$ = 65.39, $p < 0.001$). Very few people made \agae{across group - across element} comparisons. This agrees with our hypothesis that viewers are more likely to compare spatially aligned bars.

Overlaid arrangements were most likely to trigger an \agwe{across group - within element} comparison (first row, $p < 0.001$), but are least likely to trigger a \wgae{within group - across element} comparison (third row, $p < 0.001$). Adjacent arrangements were most likely to trigger a \wgae{within-group - across element} comparison (third row, $p < 0.001$), and were least likely to trigger an \agwe{across group - within element} comparison (first row, $p < 0.001$). This means that people were more likely to identify the same element and compare their values across two different groups when they view bar charts in the overlaid arrangements, and are more likely to focus on one group and compare elements within that group when they view adjacent charts.


\subsection{Visual Arrangements and Comparison Categories}
We examined how visual arrangements impacted the likelihood of participants making each of the twelve types of comparisons via a Chi-square analysis and found a significant effect (${\chi}^2$ = 132.25, $p < 0.001$). We visualize the number of comparisons participants made in each of the 12 comparisons in Figure \ref{heatmap_12categories}. Post-hoc analysis with Bonferroni's correction revealed that some visual arrangements particularly elicit certain comparison types. As summarized in Table \ref{exp2design}, overlaid arrangements especially afford C3 (all element, \agwe{across group - within element}, p = 0.001) and C4 comparisons (one-to-multiple, \agwe{across group - within element}, p = 0.037). Vertical arrangements afford type 9 comparisons (one-to-one, \wgae{within group - across element}, p = 0.017). Adjacent arrangements afford C12 comparisons (one-to-multiple, \agwe{across group - within element}, p = 0.013). Some arrangements are also particularly bad at eliciting certain comparison types. Participants were the least likely to make C3 comparisons with adjacent arrangements (p = 0.002). Vertical arrangements were the least likely to trigger C4 comparisons (p = 0.021), and overlaid arrangements were the least likely to trigger C12 comparisons (p = 0.028).

We share some participant drawings in Figure \ref{drawingComplexityExample}. You can see that the amount of effort the participant put into visually representing their comparisons differed between comparison type and arrangements, which also reflects the differing comparison affordances of the visual arrangements. For overlaid arrangements, visually annotating comparison type C3 was simple, whereas visually annotating the same comparison in an adjacent arrangement was much more complex. This corroborates with our finding that participants were more likely to make C3 comparisons when viewing overlaid arrangements, and less likely to do so when viewing adjacent arrangements. Surprisingly, although type C1 and C11 comparisons were among the most frequently made comparisons, we did not find any difference in the likelihoods of participants making them between the four visual arrangements. 


\subsection{Discussion}
\label{exp1Discussion}

Overall, as we hypothesized, viewers were more likely to compare bars that are visually aligned and spatially proximate. Vertical and adjacent arrangements best afford comparisons that involve comparing one element in one group to another element in the same group (C9: \icon{9} and C12: \icon{12}). Overlaid arrangements best afford comparisons that involve comparing one bar to all other bars in the chart (which are most often superlative comparisons, C4: \icon{4}), and comparisons that involve comparing one element to other elements, considering both groups (C3: \icon{3}). Although comparisons that involve comparing one group to another holistically (C11: \icon{11}) and comparisons that involve comparing the same element across two different groups (C1: \icon{1}) are popular comparisons, no particular visual arrangements especially afford these comparisons.

Considering that the above six comparison types (C1, C3, C4, C9, C11, C12) were commonly observed and especially afforded by the four arrangements we tested, we present them as comparison intents to visualization experts in Experiment 2 and ask them to select an arrangement that would best afford these comparison intents. There is a subtle difference between C4 and C12, as C4 is about a global extrema (viewer needs to consider all bars), while C12 is a local extrema (viewer needs to select one group and identify an extrema in it). Because both are about identifying extrema, in respect to the experts' time and to avoid a combinatorial explosion of experimental conditions, we condensed the experiment to only include C4, as finding global extremum tends to be the more common task used in other visualization evaluations (e.g., \cite{lee2016vlat}, \cite{saket2018task}).

\section{Experiment 2 Expert Intuitions}
In Study 2, we showed data visualization experts bar charts in the four different visual arrangements and asked them which one they would choose to facilitate a specific type of comparison. 

\begin{figure*}[h!]
\centering
 \includegraphics[width = \linewidth]{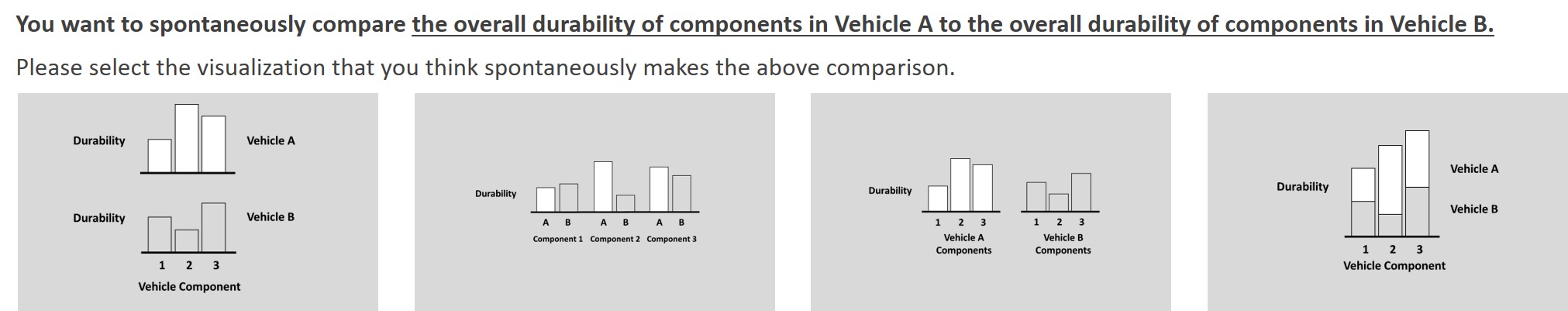}
 \caption{Multiple choice study design for experts to indicate their preferences in Experiment 2.}
 \label{exp2Example}
\vspace{-1em}
\end{figure*}

\subsection{Participant and Procedure}
We recruited 45 visualization expert volunteers from Adobe and Tableau ($M_{age} = 37.88$, $SD_{age} = 13.14$, 16 women) to complete an online survey through Qualtrics \cite{qualtrics2013qualtrics}. They reported their experience with visualizations in multiple-choice, multiple-answer questions. Among our participants, for those that chose to complete the demographic questionnaire at the end of the survey, 22 people stated that they were data analysts who used visualizations frequently or were visualization researchers, 7 people said to have taken at least one visualization design course, and 3 people indicated to be engaged with data visualization design and development (e.g., engineering, graphic design, product manager). 31\% of our participants mentioned that they enjoyed learning about visualizations through popular media and infographics. The participants also completed a subjective graph literacy report \cite{garcia2013communicating} and reported an average value of 4.75 out 6 (SD = 0.75, 1 = not good at all, 6 = extremely good), suggesting that most participants were comfortable interpreting visualizations.

Experts were given a comparison goal and asked to select the visualization they thought best makes that comparison from four arrangements via a multiple-choice task, as shown in Figure \ref{exp2Example}. 
Everyone viewed five sets of data presented with five different scenarios (e.g., Figure \ref{contextExample}), and each scenario came with one of the five listed comparison goals from Table \ref{exp2design}. The order in which the five scenarios were presented and the mapping between datasets and scenarios follow a 5 x 5 Graeco Latin Square design, such that the order in which the scenarios were presented, as well as how the datasets mapped onto the scenarios, were counterbalanced.  


\subsection{Results}
We conducted a Chi-Square analysis to investigate whether experts associate certain comparison types with certain visual arrangements. We found a significant relationship between experts' preferred visual arrangement for each comparison goal (${\chi}^2$ = 163.67, $p < 0.001$). We summarize the visual arrangements the experts identified to facilitate each comparison type in Table \ref{exp2design} and compared their intuitions to our empirical results from Experiment 1. The rightmost column shows the distribution of the experts response. From left to right, the bars represents the number of experts that selected adjacent, overlaid, stacked, and vertical arrangement as the most effective design for the given comparison goal. We see that for C4 and C9, most experts agree that overlaid is the best arrangement, but for the other three, even the experts do not agree on which arrangement might be the most effective.

Post-hoc comparisons with Bonferroni's correction suggest that experts preferred the stacked arrangement ($p < 0.001$) to make comparison C1, in contrast to crowdworkers, who collectively suggested that all four arrangements were equally likely to elicit this comparison type. For comparison C3, experts preferred the stacked arrangement ($p < 0.001$), while overlaid arrangement worked the most effectively with crowdworkers. For comparison C4, experts chose the overlaid arrangement as the most effective one (p = 0.022), which is consistent with crowdsourced results from Experiment 1. For comparison C9, experts preferred the overlaid arrangement ($p < 0.001$), but crowd-sourced results suggest that the vertical arrangement most affords this comparison type. For comparison C11, crowdsourced data from Experiment 1 suggests that the four arrangements were equally likely to elicit this comparison and experts agreed ($p > 0.1$).

\section{Design Guidelines}
\begin{figure}[h!]
\centering
 \includegraphics[width = \columnwidth]{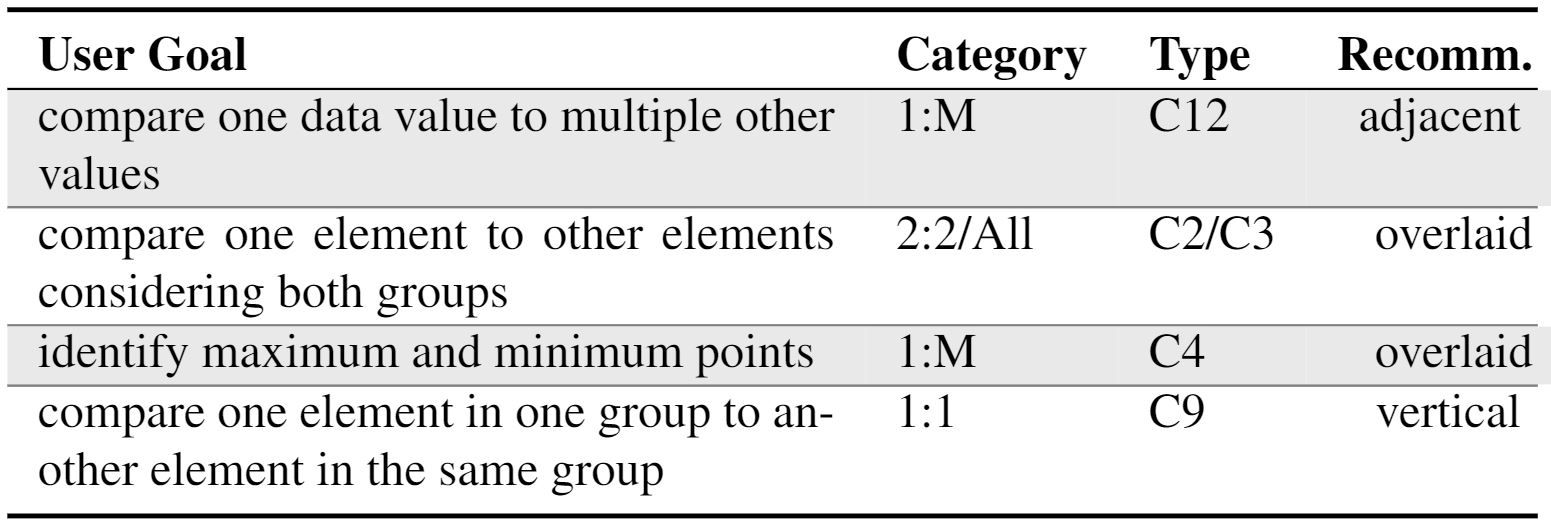}
 \caption{Recommended visual arrangements for various user goals.}
 \label{designGuidelines}
\end{figure}

We found that visual arrangements can afford different visual comparisons in bar charts, and viewers most readily compare bars that are visually aligned and spatially proximate. We recommend that visualization designers consider how the data can be best spatially arranged to facilitate key comparisons among data values.  We identify comparison affordances of the visual arrangements in Table \ref{designGuidelines}. These findings provide guidelines for a variety of visual analysis tools and applications. 
Additionally, while experts generally showed good intuitions about visualization design, there are several instances where their choices did not align with our results. This suggests that 
visualization researchers should continue to empirically explore the design affordances rather than solely relying on expert intuitions.

\noindent\textbf{Handling comparison intent in VisRec systems and NL interfaces}:  Insights from the study can be incorporated as rules for providing targeted visualization responses based on the type of comparison that the user may find useful or helps answer their question. For example, an NL utterance, ``Are paper products doing better in the West region or East region?'' is a common type of analytical inquiry. Showing a vertical arrangement of bar charts for instance, could help facilitate an effective takeaway that satisfies the user's intent.

\noindent\textbf{Smarter defaults in authoring tools}: To improve the efficacy of chart-caption pairs for visual comparisons, authors could (1) design the chart with a visual arrangement that supports the comparison goal and (2) provide a caption that emphasizes the type of comparison that the arrangement affords, beyond the current practice of just describing the variables depicted in the chart. Visual analysis tools can suggest reasonable defaults and design choices to guide the author in creating such effective chart-caption comparison pairs to doubly emphasize the comparative features in the takeaways~\cite{kim2021towards}.

\section{Limitation and Future Directions}
\label{limitsFuture}
We identify several limitations in our study that provide promising future research directions. 

First, our investigation was limited to bar charts that show groups of discrete variables. Although we experimented with different datasets and scenarios, the underlying data is relatively simple. Future research can experiment with more complex bar charts, additional datasets, and other visualization chart types. This can also lead to investigations of generalizability between visual arrangements and comparison affordances across different charts, data values, and data complexity.

Second, to address the ambiguous nature of human language, we solicited accompanying drawings that provide more detail to the sentence takeaways. As a result, we could not automate the data analysis process, and the authors had to manually read, segment, and categorize each sentence takeaway. 
We analyzed conjunction sentences by breaking them into separate comparisons. But people could make more complex comparisons involving conjunctions, usually in the form of a comparison from the results of an existing comparison. Future research could explore the visual affordances for conjunction comparisons in bar charts, and investigate ways to more effectively collect viewer takeaways via improved natural language interfaces that can automatically map verbal chart takeaways to visual comparisons and allow for a wider range of user input queries.

Third, although we found that certain visualization arrangements better afford certain visual comparisons, it is unclear whether these arrangements would also increase the accuracy of value comparisons. There might be a mismatch between what people intuitively compare in an arrangement and how accurately they can make that comparison. Future research could investigate the effectiveness of different visual arrangements from a perception angle using psychophysical methods. 

Finally, we only looked at how visual arrangements could affect how people compare elements and groups. There may be other factors that could strengthen a chart's comparison affordances. For example, highlighting aspects of a visualization has been shown to help elicit takeaways \cite{hearst2019toward, ajani2021declutter}. Future work could further explore techniques to help designers choose the best arrangement that ensures that a viewer sees the `right' story in a dataset.


\newpage
\bibliographystyle{abbrv}
\bibliography{reference}

\begin{thebibliography}{10}

\bibitem{ibmwatson}
{IBM} {W}atson {A}nalytics.
\newblock \url{http://www.ibm.com/analytics/watson-analytics/}.

\bibitem{powerbi}
{M}icrosoft {Q} \& {A}.
\newblock
  \url{https://powerbi.microsoft.com/en-us/documentation/powerbi-service-q-and-a/},
  2020.

\bibitem{thoughtspot}
{T}hought{S}pot.
\newblock \url{http://www.thoughtspot.com/}, 2020.

\bibitem{ahrens}
J.~{Ahrens}, K.~{Heitmann}, M.~{Petersen}, J.~{Woodring}, S.~{Williams},
  P.~{Fasel}, C.~{Ahrens}, C.~{Hsu}, and B.~{Geveci}.
\newblock Verifying scientific simulations via comparative and quantitative
  visualization.
\newblock {\em IEEE Computer Graphics and Applications}, 30(6):16--28, 2010.

\bibitem{ajani2021declutter}
K.~Ajani, E.~Lee, C.~Xiong, C.~N. Knaflic, W.~Kemper, and S.~Franconeri.
\newblock Declutter and focus: Empirically evaluating design guidelines for
  effective data communication.
\newblock {\em IEEE Transactions on Visualization \& Computer Graphics},
  (01):1--1, 2021.

\bibitem{alper2013}
B.~Alper, B.~Bach, N.~Henry~Riche, T.~Isenberg, and J.-D. Fekete.
\newblock Weighted graph comparison techniques for brain connectivity analysis.
\newblock In {\em Proceedings of the SIGCHI Conference on Human Factors in
  Computing Systems}, CHI 2013, pages 483 -- 492, New York, NY, USA, 2013.
  Association for Computing Machinery.

\bibitem{bakhshandeh-allen-2015-semantic}
O.~Bakhshandeh and J.~Allen.
\newblock Semantic framework for comparison structures in natural language.
\newblock In {\em Proc. of the 2015 Conference on Empirical Methods in Natural
  Language Processing}, pages 993--1002, Lisbon, Portugal, Sept. 2015.
  Association for Computational Linguistics.

\bibitem{Bierwisch1989TheSO}
M.~Bierwisch.
\newblock The semantics of gradation.
\newblock 1989.

\bibitem{clark1972process}
H.~H. Clark and W.~G. Chase.
\newblock On the process of comparing sentences against pictures.
\newblock {\em Cognitive psychology}, 3(3):472--517, 1972.

\bibitem{collins2018guidance}
C.~Collins, N.~Andrienko, T.~Schreck, J.~Yang, J.~Choo, U.~Engelke, A.~Jena,
  and T.~Dwyer.
\newblock Guidance in the human--machine analytics process.
\newblock {\em Visual Informatics}, 2(3):166--180, 2018.

\bibitem{conati2014evaluating}
C.~Conati, G.~Carenini, E.~Hoque, B.~Steichen, and D.~Toker.
\newblock Evaluating the impact of user characteristics and different layouts
  on an interactive visualization for decision making.
\newblock In {\em Computer Graphics Forum}, volume~33, pages 371--380. Wiley
  Online Library, 2014.

\bibitem{CRESSWELL1976261}
M.~CRESSWELL.
\newblock The semantics of degree.
\newblock In B.~H. PARTEE, editor, {\em Montague Grammar}, pages 261--292.
  Academic Press, 1976.

\bibitem{datasite}
Z.~Cui, S.~K. Badam, A.~Yal{\c{c}}in, and N.~Elmqvist.
\newblock Datasite: Proactive visual data exploration with computation of
  insight-based recommendations.
\newblock {\em CoRR}, abs/1802.08621, 2018.

\bibitem{scagexplorer}
T.~N. {Dang} and L.~{Wilkinson}.
\newblock Scagexplorer: Exploring scatterplots by their scagnostics.
\newblock In {\em 2014 IEEE Pacific Visualization Symposium}, pages 73--80,
  2014.

\bibitem{demiralp2017foresight}
{\c{C}}.~Demiralp, P.~J. Haas, S.~Parthasarathy, and T.~Pedapati.
\newblock Foresight: Rapid data exploration through guideposts.
\newblock {\em arXiv preprint arXiv:1709.10513}, 2017.

\bibitem{franconeri2013nature}
S.~L. Franconeri.
\newblock The nature and status of visual resources.
\newblock 2013.

\bibitem{franconeri2021CurrDirs}
S.~L. Franconeri.
\newblock Three perceptual tools for seeing and understanding visualized data.
\newblock {\em Current Directions in Psychological Science}, in press.

\bibitem{franconeri2012flexible}
S.~L. Franconeri, J.~M. Scimeca, J.~C. Roth, S.~A. Helseth, and L.~E. Kahn.
\newblock Flexible visual processing of spatial relationships.
\newblock {\em Cognition}, 122(2):210--227, 2012.

\bibitem{datatone}
T.~Gao, M.~Dontcheva, E.~Adar, Z.~Liu, and K.~G. Karahalios.
\newblock Datatone: Managing ambiguity in natural language interfaces for data
  visualization.
\newblock In {\em Proceedings of the 28th Annual ACM Symposium on User
  Interface Software Technology}, UIST 2015, pages 489--500, New York, NY, USA,
  2015. ACM.

\bibitem{garcia2013communicating}
R.~Garcia-Retamero and E.~T. Cokely.
\newblock Communicating health risks with visual aids.
\newblock {\em Current Directions in Psychological Science}, 22(5):392--399,
  2013.

\bibitem{garcia2009communicating}
R.~Garcia-Retamero and M.~Galesic.
\newblock Communicating treatment risk reduction to people with low numeracy
  skills: a cross-cultural comparison.
\newblock {\em American journal of public health}, 99(12):2196--2202, 2009.

\bibitem{gleich2010ambiguity}
B.~Gleich, O.~Creighton, and L.~Kof.
\newblock Ambiguity detection: Towards a tool explaining ambiguity sources.
\newblock In {\em International Working Conference on Requirements Engineering:
  Foundation for Software Quality}, pages 218--232. Springer, 2010.

\bibitem{Gleicher2011}
M.~Gleicher, D.~Albers, R.~Walker, I.~Jusufi, C.~D. Hansen, and J.~C. Roberts.
\newblock Visual comparison for information visualization.
\newblock {\em Information Visualization}, 10(4):289 -- 309, Oct. 2011.

\bibitem{gleitman2007give}
L.~R. Gleitman, D.~January, R.~Nappa, and J.~C. Trueswell.
\newblock On the give and take between event apprehension and utterance
  formulation.
\newblock {\em Journal of memory and language}, 57(4):544--569, 2007.

\bibitem{grahamkennedy2007}
M.~Graham and J.~Kennedy.
\newblock Exploring multiple trees through dag representations.
\newblock {\em IEEE transactions on visualization and computer graphics},
  13:1294--301, 11 2007.

\bibitem{Hamann2005COMPARINGST}
C.~Hamann, I.~Heim, D.~Lewis, P.~Seuren, and W.~Sternefeld.
\newblock Comparing semantic theories of comparison arnim von stechow.
\newblock 2005.

\bibitem{hawley2008impact}
S.~T. Hawley, B.~Zikmund-Fisher, P.~Ubel, A.~Jancovic, T.~Lucas, and
  A.~Fagerlin.
\newblock The impact of the format of graphical presentation on health-related
  knowledge and treatment choices.
\newblock {\em Patient education and counseling}, 73(3):448--455, 2008.

\bibitem{hearst2019toward}
M.~Hearst, M.~Tory, and V.~Setlur.
\newblock Toward interface defaults for vague modifiers in natural language
  interfaces for visual analysis.
\newblock In {\em 2019 IEEE Visualization Conference (VIS)}, pages 21--25.
  IEEE, 2019.

\bibitem{JardineProxies}
N.~{Jardine}, B.~D. {Ondov}, N.~{Elmqvist}, and S.~{Franconeri}.
\newblock The perceptual proxies of visual comparison.
\newblock {\em IEEE Transactions on Visualization and Computer Graphics},
  26(1):1012--1021, 2020.

\bibitem{kaplan2010lexical}
J.~Kaplan, D.~G. Fisher, and N.~T. Rogness.
\newblock Lexical ambiguity in statistics: how students use and define the
  words: association, average, confidence, random and spread.
\newblock {\em Journal of Statistics Education}, 18(2), 2010.

\bibitem{kay2016ish}
M.~Kay, T.~Kola, J.~R. Hullman, and S.~A. Munson.
\newblock When (ish) is my bus? user-centered visualizations of uncertainty in
  everyday, mobile predictive systems.
\newblock In {\em Proc. of the 2016 CHI}, pages 5092--5103, 2016.

\bibitem{kennedy:1997}
C.~Kennedy.
\newblock Projecting the adjective: The syntax and semantics of gradability and
  comparison.
\newblock 01 1997.

\bibitem{kennedy:2004}
C.~Kennedy.
\newblock Comparatives, semantics of.
\newblock {\em Encyclopedia of Language and Linguistics}, 08 2004.

\bibitem{kim2021towards}
D.~H. Kim, V.~Setlur, and M.~Agrawala.
\newblock Towards understanding how readers integrate charts and captions: A
  case study with line charts.
\newblock {\em arXiv preprint arXiv:2101.08235}, 2021.

\bibitem{Klein1980-KLEASF}
E.~Klein.
\newblock A semantics for positive and comparative adjectives.
\newblock {\em Linguistics and Philosophy}, 4(1):1--45, 1980.

\bibitem{korenjak2008clustering}
S.~Korenjak-Cerne, N.~Kej{\v{z}}ar, and V.~Batagelj.
\newblock Clustering of population pyramids.
\newblock {\em Informatica}, 32(2), 2008.

\bibitem{Larsen1978SizeSI}
A.~Larsen and C.~Bundesen.
\newblock Size scaling in visual pattern recognition.
\newblock {\em Journal of experimental psychology. Human perception and
  performance}, 4 1:1--20, 1978.

\bibitem{larsen:1998}
A.~Larsen and C.~Bundesen.
\newblock Effects of spatial separation in visual pattern matching: Evidence on
  the role of mental translation.
\newblock {\em Journal of experimental psychology. Human perception and
  performance}, 24:719--31, 07 1998.

\bibitem{Lee:2019}
D.~J.-L. Lee, H.~Dev, H.~Hu, H.~Elmeleegy, and A.~Parameswaran.
\newblock Avoiding drill-down fallacies with vispilot: Assisted exploration of
  data subsets.
\newblock In {\em Proc. of the 24th International Conference on Intelligent
  User Interfaces}, IUI '19, pages 186--196, New York, NY, USA, 2019. ACM.

\bibitem{lee2019scattersearch}
D.~J.-L. Lee, J.~Kim, R.~Wang, and A.~Parameswaran.
\newblock Scattersearch: Visual querying of scatterplot visualizations, 2019.

\bibitem{lee2021deconstructing}
D.~J.-L. Lee, V.~Setlur, M.~Tory, K.~Karahalios, and A.~Parameswaran.
\newblock Deconstructing categorization in visualization recommendation: A
  taxonomy and comparative study, 2021.

\bibitem{lee2016vlat}
S.~Lee, S.-H. Kim, and B.~C. Kwon.
\newblock Vlat: Development of a visualization literacy assessment test.
\newblock {\em IEEE transactions on visualization and computer graphics},
  23(1):551--560, 2016.

\bibitem{livingston2011}
M.~A. {Livingston} and J.~W. {Decker}.
\newblock Evaluation of trend localization with multi-variate visualizations.
\newblock {\em IEEE Transactions on Visualization and Computer Graphics},
  17(12):2053--2062, 2011.

\bibitem{Mackinlay1986}
J.~Mackinlay.
\newblock {Automating the design of graphical presentations of relational
  information}.
\newblock {\em ACM Transactions on Graphics}, 5(2):110--141, 1986.

\bibitem{Mackinlay2007}
J.~D. Mackinlay, P.~Hanrahan, and C.~Stolte.
\newblock {Show Me: Automatic presentation for visual analysis}.
\newblock {\em IEEE Transactions on Visualization and Computer Graphics},
  13(6):1137--1144, 2007.

\bibitem{matlen2020spatial}
B.~J. Matlen, D.~Gentner, and S.~L. Franconeri.
\newblock Spatial alignment facilitates visual comparison.
\newblock {\em Journal of Experimental Psychology: Human Perception and
  Performance}, 46(5):443, 2020.

\bibitem{michal2017visual}
A.~L. Michal and S.~L. Franconeri.
\newblock Visual routines are associated with specific graph interpretations.
\newblock {\em Cognitive Research: Principles and Implications}, 2(1):1--10,
  2017.

\bibitem{michal2016visual}
A.~L. Michal, D.~Uttal, P.~Shah, and S.~L. Franconeri.
\newblock Visual routines for extracting magnitude relations.
\newblock {\em Psychonomic bulletin \& review}, 23(6):1802--1809, 2016.

\bibitem{nothelfer2019measures}
C.~Nothelfer and S.~Franconeri.
\newblock Measures of the benefit of direct encoding of data deltas for data
  pair relation perception.
\newblock {\em IEEE transactions on visualization and computer graphics},
  26(1):311--320, 2019.

\bibitem{ondov2018face}
B.~Ondov, N.~Jardine, N.~Elmqvist, and S.~Franconeri.
\newblock Face to face: Evaluating visual comparison.
\newblock {\em IEEE transactions on visualization and computer graphics},
  25(1):861--871, 2018.

\bibitem{Post1995}
H.-G. Pagendarm and F.~Post.
\newblock Comparative visualization - approaches and examples.
\newblock 01 1995.

\bibitem{palan2018prolific}
S.~Palan and C.~Schitter.
\newblock Prolific. a subject pool for online experiments.
\newblock {\em Journal of Behavioral and Experimental Finance}, 17:22--27,
  2018.

\bibitem{qu2017keeping}
Z.~Qu and J.~Hullman.
\newblock Keeping multiple views consistent: Constraints, validations, and
  exceptions in visualization authoring.
\newblock {\em IEEE transactions on visualization and computer graphics},
  24(1):468--477, 2017.

\bibitem{qualtrics2013qualtrics}
I.~Qualtrics.
\newblock Qualtrics.
\newblock {\em Provo, UT, USA}, 2013.

\bibitem{rensink2002}
R.~A. Rensink.
\newblock Change detection.
\newblock {\em Annual Review of Psychology}, 53(1):245--277, 2002.
\newblock PMID: 11752486.

\bibitem{roberts2007}
J.~C. {Roberts}.
\newblock State of the art: Coordinated multiple views in exploratory
  visualization.
\newblock In {\em CMV 2007}, pages 61--71, 2007.

\bibitem{roth2012asymmetric}
J.~Roth and S.~Franconeri.
\newblock Asymmetric coding of categorical spatial relations in both language
  and vision.
\newblock {\em Frontiers in psychology}, 3:464, 2012.

\bibitem{saket2018task}
B.~Saket, A.~Endert, and {\c{C}}.~Demiralp.
\newblock Task-based effectiveness of basic visualizations.
\newblock {\em IEEE transactions on visualization and computer graphics},
  25(7):2505--2512, 2018.

\bibitem{Sapir1944GradingAS}
E.~Sapir.
\newblock Grading, a study in semantics.
\newblock {\em Philosophy of Science}, 11:93 -- 116, 1944.

\bibitem{Schwarzchild2002QuantifiersIC}
R.~Schwarzchild and K.~Wilkinson.
\newblock Quantifiers in comparatives: A semantics of degree based on
  intervals.
\newblock {\em Natural Language Semantics}, 10:1--41, 2002.

\bibitem{eviza}
V.~Setlur, S.~E. Battersby, M.~Tory, R.~Gossweiler, and A.~X. Chang.
\newblock Eviza: A natural language interface for visual analysis.
\newblock In {\em Proc. of the 29th Annual Symposium on User Interface Software
  and Technology}, UIST '16, pages 365 -- 377, New York, NY, USA, 2016.
  Association for Computing Machinery.

\bibitem{Setlur:IUI}
V.~Setlur, M.~Tory, and A.~Djalali.
\newblock Inferencing underspecified natural language utterances in visual
  analysis.
\newblock In {\em Proc. of the 24th International Conference on Intelligent
  User Interfaces}, IUI '19, pages 40--51, New York, NY, USA, 2019. Association
  for Computing Machinery.

\bibitem{shah2011bar}
P.~Shah and E.~G. Freedman.
\newblock Bar and line graph comprehension: An interaction of top-down and
  bottom-up processes.
\newblock {\em Topics in cognitive science}, 3(3):560--578, 2011.

\bibitem{srinivasan2018s}
A.~Srinivasan, M.~Brehmer, B.~Lee, and S.~M. Drucker.
\newblock What's the difference? evaluating variations of multi-series bar
  charts for visual comparison tasks.
\newblock In {\em Proceedings of the 2018 CHI Conference on Human Factors in
  Computing Systems}, pages 1--12, 2018.

\bibitem{Tait2010}
A.~R. Tait, T.~Voepel-Lewis, B.~J. Zikmund-Fisher, and A.~Fagerlin.
\newblock Presenting research risks and benefits to parents: does format
  matter?
\newblock {\em Anesthesia and analgesia}, 111(3):718, 2010.

\bibitem{tufte_envisioning_1990}
E.~R. Tufte.
\newblock {\em Envisioning information}.
\newblock Graphics Press, Cheshire, Conn., 1990.

\bibitem{tversky2014visualizing}
B.~Tversky.
\newblock Visualizing thought.
\newblock In {\em Handbook of human centric visualization}, pages 3--40.
  Springer, 2014.

\bibitem{wongsuphasawat2016towards}
K.~Wongsuphasawat, D.~Moritz, A.~Anand, J.~Mackinlay, B.~Howe, and J.~Heer.
\newblock Towards a general-purpose query language for visualization
  recommendation.
\newblock In {\em Proc. of the Workshop on Human-In-the-Loop Data Analytics},
  page~4. ACM, 2016.

\bibitem{wu2021survey}
A.~Wu, Y.~Wang, X.~Shu, D.~Moritz, W.~Cui, H.~Zhang, D.~Zhang, and H.~Qu.
\newblock Survey on artificial intelligence approaches for visualization data.
\newblock {\em arXiv preprint arXiv:2102.01330}, 2021.

\bibitem{xiong2019illusion}
C.~Xiong, J.~Shapiro, J.~Hullman, and S.~Franconeri.
\newblock Illusion of causality in visualized data.
\newblock {\em IEEE Transactions on Visualization and Computer Graphics},
  26(1):853--862, 2019.

\bibitem{xiong2021visual}
C.~Xiong, C.~Stokes, and S.~Franconeri.
\newblock Visual salience and grouping cues guide relation perception in visual
  data displays.
\newblock {\em Journal of Vision}, 2021.

\bibitem{xiong2019curse}
C.~Xiong, L.~van Weelden, and S.~Franconeri.
\newblock The curse of knowledge in visual data communication.
\newblock {\em IEEE transactions on visualization and computer graphics}, 2019.

\bibitem{xu2015capacity}
Y.~Xu and S.~L. Franconeri.
\newblock Capacity for visual features in mental rotation.
\newblock {\em Psychological science}, 26(8):1241--1251, 2015.

\bibitem{zacks1999bars}
J.~Zacks and B.~Tversky.
\newblock Bars and lines: A study of graphic communication.
\newblock {\em Memory \& Cognition}, 27(6):1073--1079, 1999.

\bibitem{zhu2020survey}
S.~Zhu, G.~Sun, Q.~Jiang, M.~Zha, and R.~Liang.
\newblock A survey on automatic infographics and visualization recommendations.
\newblock {\em Visual Informatics}, 4(3):24--40, 2020.

\end{thebibliography}

\end{document}